# Safeguarding Skies: Airport Cybersecurity in the Digital Age

Suphannee Sivakorn*, Nuttaya Rujiratanapat*, Yotsapat Ruangpaisarn*,
Chanond Duangpayap* and Sakulchai Saramat*

* Corresponding Author: Suphannee Sivakorn, E-mail: suphannee_si@rmutto.ac.th

**Abstract**

The aviation industry faces significant vulnerabilities from both physical and cybersecurity threats, highlighting the urgent need for enhanced cybersecurity measures amid increasingly sophisticated attacks. This paper systematically reviews emerging threats at airports, analyzing real-world incidents and relevant literature while mapping risks to the MITRE ATT&CK Matrix, a widely recognized knowledge base for categorizing cyberattack tactics, techniques, and procedures. This is the first to apply the MITRE Matrix to airport security risks, offering a novel approach to understanding and mitigating these challenges. Building on this analysis, the paper advocates for modern cybersecurity defense models, emphasizing Cybersecurity Frameworks and Zero Trust Architecture, as well as critical measures for supply chain risk management and strategies to mitigate ransomware and DoS attacks. Our analysis provides insights into vulnerabilities and actionable recommendations, serving as a comprehensive guide for aviation stakeholders to strengthen defenses against evolving cybersecurity threats.

## 1. Introduction

In the physical domain, airport security entails the screening of passengers, baggage, cargo, and the fortification of secure areas within the airport premises. Airport authorities undertake substantial efforts to prevent unlawful interference and ensure their security practices meet current standards. However, in the realm of technology, cybersecurity often receives inadequate attention than physical security, as evidenced by a 2017 survey of the top major airports in Europe and the U.S., wherein only 59% of respondents claimed to have an effective cybersecurity policy [1]. This oversight is concerning, especially given a 530% increase in cyberattacks within the aviation industry from 2019 to 2022 [2]. Recent initiatives by authorities, including the Transportation Security Administration (TSA) and the International Air Transport Association (IATA), emphasize the need for enhanced cybersecurity measures, mandating proactive steps to mitigate cyber threats [3], [4].

In an effort to strengthen the cybersecurity posture of the aviation industry, this study comprehensively explores existing literature and recent data on airport cybersecurity. We examine airport technologies, security concerns, and recent cyber incidents in Section 2, and outline our research methodology in Section 3. Section 4 presents a systematic literature review of airport cybersecurity from the past five years, and Section 5 categorizes the current landscape and identifies relevant risks. We map these risks to the MITRE ATT&CK Matrix for Enterprise, a widely recognized cybersecurity knowledge base for developing effective security strategies in Section 6. Section 7 presents modern cybersecurity defense strategies, advocating for Cybersecurity Frameworks and Zero Trust Architecture and highlights essential measures for mitigating supply chain risks, ransomware, and denial-of-service (DoS) attack. Section 8 discusses key challenges

*Department of Computer Science, Faculty of Science and Technology, Rajamangala University of Technology Tawan-ok*

and outlines future research directions in this field, with the conclusion presented in Section 9.

The major contributions of this paper are as follows:

- We conducted an extensive analysis of recent cyberattacks and a literature review from the past five years, categorizing key cybersecurity risks into nine distinct areas to clarify the challenges faced by modern airport operations.
- We correlate these identified risks with the MITRE ATT&CK Matrix for Enterprise, making this paper the first to map airport security risks to the Matrix. This serves as a valuable tool for exploring each risk through practical defenses and best practices outlined in the MITRE knowledge base.
- Based on our analysis, we advocate for adopting modern security practices, including Cybersecurity Frameworks and Zero Trust Architecture, along with practical measures to defend against evolving airport cyber threats.

## 2. Background

Understanding airport technologies, threat actors, and recent high-profile incidents highlights the need for stronger security measures. This section examines airport technology, cybersecurity concerns, the landscape of cyber threat actors, and notable attack incidents.

### 2.1 Airport Technology and Security Concerns

Airport operations have transformed significantly to support the global aviation industry growth, leading to advancements in technology aimed at enhancing efficiency and service. The evolution of airport technology is delineated into four stages: Airport 1.0 - 4.0.

**Airport 1.0** primarily focuses on ensuring the physical safety of operations, with no security concerns [5].

**Airport 2.0** incorporates technologies for collaboration technologies such as IP telephony, broadband, and Wi-Fi [6]. Following the events of 9/11, the TSA was established to oversee transport security [7].

**Airport 3.0 or "Smart Airport"** integrates the Internet of Things (IoT), Artificial Intelligence, smart sensors to enhance passenger experience [8].

**Airport 4.0** emphasizes the use of technologies to support airport operations and enhance passenger experiences, with a focus on data analytics as a core capability [9], [10].

While these airport advancements offer notable benefits, they are susceptible to interference and malicious modification without proper deployment.

### 2.2 Airport Cyber Threat Actors

Incidents of air terrorism have led adversaries to adapt their tactics in both physical and cyber domains [7]. This section outlines four types of cyber threat actors:

**1: Advanced Persistent Threat (APT):** Organized groups or foreign governments motivated by political or economic goals. They often target critical infrastructure, including airports [11] - [16].

**2: Cybercrime:** Attackers in this category target systems for valuable and sensitive information from passengers and airport employees [17] - [21], particularly those that are internet-facing or publicly accessible [17].

**3: Peer Group Service Disruption:** Hackers motivated by political agendas or beliefs whose focus is on service disruptions rather than data theft and financial gain [22] - [30].

**4: Insider Threats:** Risks that originate from within the organization, typically associated with current or former members of the organization, it may also arise from third parties such as contractors and temporary workers.

### 2.3 Recent Notable Cybersecurity Incidents

The urgency of cybersecurity in airports has become apparent through numerous studies [1], [5], [8]. From 2022 to 2024, various notable incidents have highlighted cyberattacks affecting airport operations and public perception. Table 1 details and categorizes these incidents by attack type, including Denial-of-Service, Ransomware, Vulnerability Exploitation, and Phishing. This analysis will assist in identifying cybersecurity risks and associated attack vectors for airports in Section 5.

**Denial-of-Service (DoS).** Recently, several major U.S. airports were targeted by coordinated DoS attacks [29], which overloaded airport servers. Similar incidents have occurred at various airports worldwide [23] - [31]. In some cases, attackers have demanded cryptocurrency payments to stop the attacks, exploiting the difficulty of tracing such transactions [32].

**Ransomware.** The airport industry has experienced a significant increase in ransomware attacks, primarily due to system vulnerabilities and phishing attempts [33] - [35], [38]. In 2024, notable incidents led to delays in passenger processing and flight schedules [34], [35], while others resulted in the leakage of sensitive data [33].

**Vulnerability Exploitation** poses significant risks for airports, which rely on variety of software applications for their operations, ranging from flight scheduling, air traffic control, baggage handling, and security systems [1], [5], [8]. This diversity broadens the attack surface, introducing potential vulnerabilities and inadequate security practices from vendor [17], [19], [20] - [22], [25], [36], [37].

**Phishing.** Although no new incidents have been disclosed recently, phishing remains a significant threat with airport employees and customers vulnerable to scams [38]. During a recent global outage linked to CrowdStrike [39], opportunistic hackers exploited the situation by sending fake information to scam IT personnel [40].

## 3. Research Methodology

This study synthesizes recent airport cybersecurity incidents, literature, insights from online sources, and an examination of various cybersecurity standards and policies. Our goal is to identify and delineate the prevailing cybersecurity threats and risks, categorizing them in alignment with the MITRE ATT&CK Matrix. By systematically mapping these risks to the Matrix, we provide a strategic approach for mitigating cybersecurity risks and applying effective defense techniques based on current best practices.

## 4. Literature Review

We conduct a comprehensive literature review by searching academic databases, including Google Scholar, ResearchGate, Scopus, and Web of Science, using the following keywords: "airport AND cybersecurity", "aviation AND cybersecurity", "airport AND information security", "airport AND IT security", "smart airport", and "airport AND cyber risk". Our focus was on peer-reviewed studies from the last five years addressing the impacts of cybersecurity on modern airports, challenges, and risks, while excluding studies on airport physical security or unrelated aviation topics. In total, we reviewed 31 publications, categorizing them into eight primary areas: (1) Critical Infrastructure, (2) IoT, Smart Devices, and AI Technology, (3) Supply Chain, (4) Cybersecurity Awareness, (5) Risk and Threat Analysis, (6) Standards and Regulations, (7) Cybersecurity Framework, and (8) Case Studies and Surveys. Table 2 presents the number of publications in each category and highlights specific airport security risks where applicable.

**Critical Infrastructure.** This category focuses on the critical infrastructures of airports, such as communication protocols, Air Traffic Management (ATM), and surveillance technologies [41] - [47]. These studies analyze vulnerabilities and mitigations, with examples including man-in-the-middle attacks between aircraft and ground control [42], the lack of encryption in the Automatic Dependent Surveillance-Broadcast (ADS-B) [43], [45], and the security concerns related to digitization of the Traffic Collision Avoidance System (TCAS) [44]. These findings underscore the need to address cybersecurity risks in airports. To this end, we associate these publications related to airport security risks as follows: (1) Insecure Network Architecture, (2) Malware and Ransomware (3) Data Breach and (4) DoS.

**IoT, Smart Devices, and AI Technology.** Research here addresses cybersecurity risks from IoT devices and AI technologies used in airport operations [1], [5], [8], [48] - [50]. Consequently, we associate these publications with specific risks, including: (1) Public-facing Access, (2) Insecure Network Architecture, (3) Internet-facing Applications and Services,

*Table 1. Summary of Publicly Disclosed Notable Cybersecurity Incidents at Airports (2022-2024).*

| Attack Incident | Attack Incident Summary | Year | Attack Technique | Impact on Airport Services | | | Threat Actor Type* |
| --- | --- | --- | --- | --- | --- | --- | --- |
| | | | | Operational Disruption | Website or Application | Data Leakage | |
| Seattle Airport [33] | The Port of Seattle confirmed that a ransomware attack caused significant outages at Seattle-Tacoma International Airport, affecting services like Wi-Fi, check-in kiosks, and passenger displays. The attack also resulted in some data being stolen and encrypted | 2024 | Ransomware | ● | | ● | 2 |
| Pau-Pyrénées Airport [34] | Pau-Pyrénées Airport was hit by a ransomware attack from the MONTI group, which exfiltrated sensitive data and published it on the dark web. | 2024 | Ransomware | | | ● | 2 |
| Croatia's Split Airport [35] | Split Airport in Croatia experienced a ransomware attack that resulted in flight cancellations and delays. The incident has been linked to the Akira group, which is associated with the Russian-based Conti group. | 2024 | Ransomware | ● | | | 2 |
| Los Angeles International Airport [36] | A hacker group, IntelBroker, exploited the airport's CRM system vulnerability, accessing a database with sensitive information (e.g., private plane owners' full names, emails, CPA numbers) | 2024 | Vulnerability Exploitation | | | ● | 2 |
| Copenhagen Airport [31] | The airport website was taken offline. Passengers were advised to use their smartphones as an alternative to receive updates on their flights. | 2024 | DoS | | ● | | unknown |
| Beirut International Airport [17] | Hackers displayed a message on screens at the airport threatening to bomb the airport. | 2024 | unknown | ● | | | 3 |
| Long Beach Airport [18] | Part of Long Beach City system cyberattack. The website was taken offline. | 2023 | Ransomware | | ● | | 2 |
| Cairo International Airport [23] | The airport website and mobile application were taken down. Anonymous Collective hacker group took credit for the attack. | 2023 | DoS | | ● | | 3 |
| Czech and Prague Airport [24] | The airport website was taken offline. | 2023 | DoS | | ● | | 3 |
| Querétaro Intercontinental Airport [19] | LockBit ransomware hacker group took credit for the attack, threatening to leak data. The airport claimed that stolen data was in the public domain. | 2023 | Ransomware | | | ● | 2 |
| Montreal-Trudeau International Airport [25] | Border checkpoint outages e.g., check-in kiosks and electronic gates caused significant delays. A hacker group, NoName057(16) claimed responsibility. | 2023 | DoS | ● | | | 3 |
| Charles de Gaulle Airport [26] | The airport website was taken offline. Cybercriminal, Dark Storm, claimed responsibility. | 2023 | DoS | | ● | | 3 |
| UK Airports [27] | The airport website was taken offline. UserSec hacker group claimed the responsibility. | 2023 | DoS | | ● | | 3 |
| Kenya Airports Authority [20] | Data breach incident. Attackers released data including procurement plans, physical plans, site surveys, invoices and receipts. | 2023 | unknown | | | ● | 2 |
| German Airports [28] | Several German airports' websites were taken offline. | 2023 | DoS | | ● | | 3 |
| US Major Airports [29] | Coordinated DoS attacks targeted several major US airports. A hacker group, Killnet claimed responsibility. | 2022 | DoS | | ● | | 3 |
| Brazil Airports [37] | Rio de Janeiro airport's electronic displays were hacked to show pornographic movies instead of ads and flight info. | 2022 | Vulnerability Exploitation | ● | | | unknown |
| Italian Airports [30] | Coordinated DoS attacks targeted several Italian airports. A hacker group, Killnet claimed responsibility. | 2022 | DoS | | ● | | 3 |
| Swissport at Zurich Airport [21] | Airport ground services and air cargo, Swissport, were hit with a ransomware attack causing Zurich Airport operation disruptions. | 2022 | Ransomware | ● | | | 2 |

The sign ● indicates the impact on airport services from the attack.

*The Threat Actor Type number delineates the category of cyber threat actors in Section 2.2

(4) Malware and Ransomware, (5) Data Breach, and (6) DoS as these threats exploits internet connectivity used by IoT and smart devices. Furthermore, vulnerabilities in these products can lead to supply chain attacks [51].

**Supply Chain and Third Party.** This category examines cybersecurity vulnerabilities arising from supply chain and third-party partnerships. For example, Hann (2020) emphasized the complex socio-technical landscape of the ATM System [47], emphasizing the need for attention to sectors critical to airport operations, particularly in the context of digital cyber warfare [51], as discussed.

**Cybersecurity Awareness.** This category investigates the effectiveness of cybersecurity awareness training within airport environments. While numerous studies have highlighted the significance of cybersecurity awareness training [1], [5], [8]. However, only one recent publication by Sabillon et al. (2023) [52] has thoroughly examined this topic. We categorize these publications under the following risks: (1) Social Engineering, (2) Insider Threats, and (3) Data Breach, as these risks often arise from human [1], [5], [8], [52] - [54].

**Risk and Threat Analysis.** This research category conducts literature reviews to identify risks and threats affecting airports. Studies provide insights into threats and recommend improvements for threat detection and response [41], [43], [49], [55] - [64]. Numerous works study airport cybersecurity incidents [55] - [60], including threat actor typologies [58], [61] and associated risks and threats in relation to ICAO (International Civil Aviation Organization) standards [55], which encompass the entire spectrum of airport security risks.

**Standards and Regulations.** Publications in this category review existing cybersecurity standards and regulations pertinent to the aviation sectors [63] - [66]. They study challenges and gaps in airport cybersecurity policies posed by rapid technological development and call for international cooperation and standardized policies, which currently remain insufficient [64].

**Cybersecurity Framework.** This category investigates frameworks tailored for airports, focusing on models that systematically manage risks and enhance resilience. While many studies agree the necessity of these frameworks [1], [5], [8], [41], [55], [67], [68] for example, adopting the National Institute of Standards and Technology's (NIST) Cybersecurity Framework to comply with ICAO standards [55], only few recent publications [55], [67], [68] provide actionable details. Nevertheless these frameworks often lack comprehensive insights into adversary behavior, which are essential for identifying and responding to threats throughout an attack's lifecycle. Further details will be provided in Section 6.

**Case Study and Survey.** This category focuses on research examining the cybersecurity posture of specific airports. Publications may present case studies based on geography [60], [62], [63], or specific events [69], [70] like the COVID-19 pandemic [70] to gather insights on cybersecurity practices and challenges.

## 5. Airport Security Risks

This section provides detailed exploration of cybersecurity risks associated with attack vectors (Section 2.3) and those identified in our literature review (Section 4).

### 5.1 Public-facing Accesses

**BYOD.** The practice of Bring-Your-Own-Device (BYOD) commonly raises concern due to the exposure of organizations to vulnerabilities [71], [72]. However, in airport settings, passengers commonly use their personal devices. The diversity of connected devices in this context complicates device management [73], heightening the security risk when these devices connect to airport networks.

**Public Access Services** such as Wi-Fi access, check-in kiosks, and charging stations enhance the passenger experience [6], but they also increase risks by leaving users vulnerable to cyberattacks [74] - [76], particularly if proper network segmentation is not implemented.

**Man-in-the-Middle (MITM)** attacks are prevalent on public Wi-Fi, where cybercriminals eavesdrop using network snooping and sniffing tools to steal sensitive information [75], [77]–[79]. Despite this, many critical websites continue to serve content over unencrypted connections [79] - [81].

---

[1]NIST Cybersecurity Framework:
https://www.nist.gov/cyberframework

*Table 2.* Airport Cybersecurity Publications by Category and Associated Security Risks.

| Publication Category | List of Publications | Number of Publications | Associated Security Risks (when applicable) |
|---|---|---|---|
| Critical Infrastructure | [41] - [47] | 9 | Insecure Network Architecture<br>Malware and Ransomware<br>Data Breach<br>DoS |
| IoT, Smart Devices and AI Technology | [1], [5], [8], [48] - [50] | 6 | Public-facing Access<br>Insecure Network Architecture<br>Internet-facing Applications and Services<br>Malware and Ransomware<br>Data Breach<br>Supply Chain and Third Party<br>DoS |
| Supply Chain and Third Party | [47] | 1 | Supply Chain and Third Party |
| Cybersecurity Awareness | [52] | 1 | Social Engineering<br>Insider Threats<br>Data Breach |
| Risk and Threat Analysis | [41], [43], [49], [55] - [64] | 13 | Public-facing Access<br>Insecure Network Architecture<br>Internet-facing Applications and Services Social Engineering<br>Malware and Ransomware<br>Data Breach<br>Supply Chain and Third Party<br>Insider Threat<br>DoS |
| Standard and Regulation | [63] - [66] | 4 | (inapplicable) |
| Cybersecurity Framework | [55], [67], [68] | 3 | (inapplicable) |
| Case Study and Survey | [60], [62], [63], [69], [70] | 5 | (inapplicable) |
| Total | | 31 | Public-facing Access<br>Insecure Network Architecture<br>Internet-facing Applications and Services Social Engineering<br>Malware and Ransomware<br>Data Breach<br>Supply Chain and Third Party<br>Insider Threat<br>DoS |

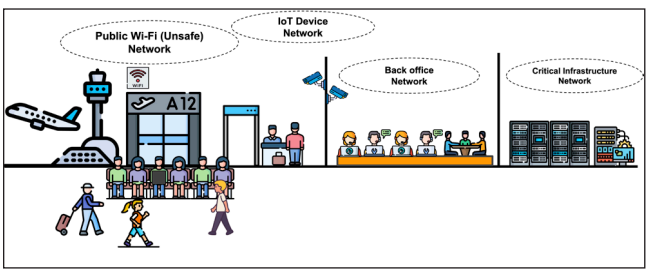

*Figure 1.* Basic Network Segmentation for Airport Security.

**Malware from Untrusted Devices.** In this attack, bad actors aim to inject malicious payload onto Wi-Fi users' devices [82], [83]. Adversaries may target vulnerabilities on popular devices, e.g., iOS [84]. Public charging stations also pose risks, known as "juice jacking", where attackers use these stations to spread malware and extract data from smartphones [85].

**Malicious Hotspots.** A malicious hotspot, also known as a "rogue access point" poses a significant threat to public Wi-Fi users. The attacker sets up a wireless access point with an identical SSID to deceive users, making users vulnerable to MITM or other network attacks [86] - [88].

### 5.2 Insecure Network Architecture

An insecure network architecture may allow attackers to gain access to internal systems and move laterally across organization assets. Effective network segmentation is a key component, enabling administrators to manage network interactions more securely by implementing security policies, with varying levels of security and trust assigned to different applications [89]. Figure 1 illustrates a basic example of network segmentation applicable to an airport, where each segment requires specific measures and is separated based on the different entities and stakeholders involved, which can be described as follow:

**Public Wi-Fi.** To prevent potential malicious activities spreading to other airport entities such as CCTV systems [1] and malware incidents at Vienna Airport [90], it should be completely segregated from other airport networks.

**IoT Devices.** IoT devices often rely on vendor or third-party-based solutions, making them vulnerable to third-party security risks (Section 5.7). Consequently, it is recommended to isolate them from other networks, particularly critical networks.

**Back Office** is responsible for the administration, operations and logistics of the airport. Given the human-centric nature of these operations, this network is prone to risks such as phishing, social engineering, and other human errors. This network should be kept separated for added security.

**Critical Infrastructures** includes crucial assets for the airport operations. Access to this network should be restricted from the public network, robust authentication and encryption measures must be implemented, as highlighted in several studies discussed in Section 4.

### 5.3 Internet-facing Applications and Services

**Security Vulnerabilities.** Adversaries often exploit weaknesses in internet-facing applications such as airport websites, and mobile applications. These vulnerabilities can arise from software bugs, design flaws, or unpatched vulnerabilities, as discussed in Section 2.3 and Section 4.

**Weak authentication** practices in internet-facing applications pose significant risks for airports, potentially leading to unauthorized access to critical systems. Additionally, remote access for employees can further complicate security; if authentication credentials are weak, attackers may gain broader access to internal networks [91].

### 5.4 Social Engineering

**Phishing.** Social engineering attacks exploit human vulnerabilities, with phishing being a significant concern. In 2013, over 75 U.S. airports reported incidents involving phishing emails designed to deceive users into disclosing financial information [61]. Some of these attacks predominantly target employees with privileged access to critical systems [8].

### 5.5 Malware and Ransomware

Ransomware incidents often lead to airport operational disruptions and passenger experience [18], [19], [21], [33] - [35]. Additionally, malware attacks may lead to data breaches, exposing sensitive information such as passenger records, payment details, and employee credentials [31], [32]. Such breaches jeopardize privacy and can incur financial costs for airports, including remediations and regulatory fines.

### 5.6 Data Breach

Data breaches often results in the unauthorized access, disclosure, or theft of sensitive information [19], [20], [33], [34], [36]. Additionally, breaches of sensitive operational information can undermine airport operations and lead to security vulnerabilities [20]. In some cases, attackers may exfiltrate data and demand ransom for its return or for the decryption of compromised systems [32].

### 5.7 Supply Chain and Third-Party

**Security Vulnerabilities** in systems can allow attackers to gain unauthorized access. These weaknesses may stem from known or unknown third-party software and hardware bugs, and misconfigurations [92]. Zero-day vulnerabilities pose particular risks, as attacks can occur before developers issue patches. Concerns about IoT and vendor solution vulnerabilities are amplified by the potential for threat actors to compromise not only the affected device but also other network assets [93].

**No Security Update Mechanism.** Many solutions, particularly IoT devices, may lack a security update mechanism, leaving them vulnerable even after patches have been released [94].

**No Common Standards and Specifications.** The lack of universally accepted standards for IoT device development leads to inconsistent implementations and design choices, which negatively affect security. Users must manage multiple technologies to effectively support these devices [5].

**Supply Chain Compromise** involves manipulating products

before they reaches consumer, creating vulnerabilities in critical systems. For example, compromised chips or drivers in smart devices at airports can expose systems to attack [95]. High-profile incidents, such as the SolarWinds hack, affected over 18,000 networks globally [96]. Additionally, a recent incident in Lebanon further illustrates the dangers, where devices were reportedly manipulated for digital warfare [51].

**No Physical Hardening.** IoT devices are often deployed in various locations throughout the airport, making them vulnerable to tampering during unattended operations. Physical access can result in theft and unauthorized access to internal circuits and overwriting changes [1].

### 5.8 Insider Threat

An insider threat is a security risk posed by individuals who misuse their access or privileged accounts. A malicious insider, often referred to as a "Turncloak," intentionally abuses legitimate access to steal sensitive information or manipulate critical aviation systems. Mitigating this threat involves adhering to information security management standards and guidelines [97].

### 5.9 Denial-of-Service

As outlined in Section 2, DoS attacks on airport websites are significant threats to the aviation sector, with recent treads showing demands ransom payments to stop these attacks, aided by the anonymity of cryptocurrencies [32].

## 6. Airport Security Risks and MITRE ATT&CK Matrix

This section provides a comprehensive analysis of the security risks faced by airports, categorizing these risks in alignment with the MITRE ATT&CK Matrix for Enterprise (or MITRE Matrix) [98]. This widely adopted cybersecurity knowledge base outlines the tactics, techniques and procedures (TTPs) utilized globally for threat analysis and security defenses. Notably, this paper is the first to propose applying the MITRE Matrix to bolster the cybersecurity posture of airports. We correlate all identified airport security risks—derived from cybersecurity incidents and a systematic literature review—with MITRE techniques to mitigate cyberattacks arising from these identified risks.

### 6.1. MITRE Matrix: TTPs

The MITRE Matrix categorizes attacker tactics and techniques. Each tactic represents a high-level goal, while the techniques describe the specific methods employed to achieve that goal, both indexed for easy reference. Each technique includes (1) procedures based on real-world incidents, (2) mitigations with actionable defense recommendations such as configurations and tools, and (3) detection strategies and recommendations for identifying the attacks. We believe that this comprehensive information enables airport security personnel to effectively implement strategies to mitigate identified risks.

### 6.2 Airport Security Risks with the MITRE Matrix

The MITRE Matrix is a valuable tool for identifying and mapping airport security risks related to potential attacks. Given the complexity of vulnerabilities, some risks may align with multiple MITRE techniques. This paper focuses on two key tactics: Initial Access (TA001) and Impact (TA0040). Initial Access is fundamental as it represents the first step for adversaries to gain entry into protected systems. The Impact tactic, particularly T1498 (Network Denial of Service), is emphasized due to its prevalence due to its frequency in recent incidents discussed in Section 2.3.

Table 3 provides an overview of categorized airport security risks and their associated MITRE techniques, listing all ten Initial Access techniques and one Impact technique (retrieved September 2024). Each technique is identified and referenced by an ID (e.g., T1189, T1190).

### 6.3 MITRE Initial Access Techniques (TA0001)

Initial Access is a critical phase in the cyber kill chain, representing the methods adversaries use to gain entry into target systems. Below are the relevant techniques from the MITRE Matrix associated with Initial Access and the corresponding airport security risks:

**Public-facing Access (T1659, T1190, T1200):** Techniques such as Content Injection (T1659) allow attackers to insert malicious content into network traffic, often through public Wi-Fi. Exploiting vulnerabilities in public-facing applications (T1190),

---

[2] MITRE Matrix Initial Access Tactic:
https://attack.mitre.org/tactics/TA0001/

[3] MITRE Matrix Impact Tactic:
https://attack.mitre.org/tactics/TA0040/

*Table 3.* Summary of Airport Security Risks Linked to MITRE Matrix Tactics and Techniques.

| Airport Security Risk | Initial Access: TA0001 | | | | | | | | | | Impact: TA0040 |
|---|---|---|---|---|---|---|---|---|---|---|---|
| | T1659 | T1189 | T1190 | T1133 | T1200 | T1566 | T1091 | T1195 | T1199 | T1078 | T1498 |
| 1. Public-facing Accesses | • | | • | | • | | | | | | |
| 2. Insecure Network Architecture | • | | • | • | | | | | | | |
| 3. Internet-facing Applications and Services | | | • | • | | | | | | | |
| 4. Social Engineering Attacks | • | • | | | | • | • | | | • | |
| 5. Malware and Ransomware | • | • | • | • | • | • | • | • | • | • | |
| 6. Data Breach | • | • | • | • | • | • | • | • | • | • | |
| 7. Supply Chain and Third Party | | | | • | • | | | • | • | | |
| 8. Insider Threats | | | | | | • | | | • | • | |
| 9. DoS | | | | | | | | | | | • |

The sign ● indicates that the airport security risk shown can be categorized according to the specific MITRE Matrix technique.

Associated MITRE Initial Access Tactic (TA0001)

| ID | Technique |
|---|---|
| T1659 | Content Injection |
| T1189 | Drive-by Compromise |
| T1190 | Exploit Public-Facing Application |
| T1133 | External Remote Services |
| T1200 | Hardware Additions |
| T1566 | Phishing |
| T1091 | Replication Through Removable Media |
| T1195 | Supply Chain Compromised |
| T1199 | Trusted Relationship |
| T1078 | Valid Accounts |

Associated MITRE Impact Tactic (TA0040)

| ID | Technique |
|---|---|
| T1498 | Network Denial of Service |

such as kiosks and charging stations, can provide unauthorized access due to unpatched vulnerabilities or misconfigurations. This may be coupled with Hardware Additions (T1200), where attackers exploit exposed ports to introduce unauthorized devices [99].

**Insecure Network Architecture (T1659, T1190, T1133):** Techniques such as Content Injection (T1659) and Exploiting Public-facing Applications (T1190) become more dangerous in poorly secured environments, allowing attackers to gain initial access and subsequently move laterally. Additionally, External Remote Services (T1133) may compromise internal network by enabling unauthorized access through insecure remote connections.

**Internet-facing Applications and Services (T1190, T1133):** The presence of internet-facing applications and services exposes airports to significant cybersecurity risks via techniques such as Exploiting Public-facing Applications (T1190) and External Remote Services (T1133). Attackers can target vulnerabilities within publicly accessible systems—like online booking websites and service APIs—to gain unauthorized access to sensitive assets. Insecure remote services can also create entry points for attackers, allowing them to gain unauthorized access to internal systems through airport VPNs [100].

**Social Engineering (T1659, T1189, T1566, T1091, T1078):** Techniques such as Content Injection (T1659), Drive-by Compromise (T1189), and Phishing (T1566) are utilized to manipulate victims. The use of insecure removable media (T1091) may allow untrusted devices to introduce malware to critical systems [101]. Technique like Valid Accounts (T1078) may enable attackers to gain access via stolen account.

**Malware, Ransomware and Data Breach (T1659, T1189, T1190, T1133, T1200, T1566, T1091, T1195, T1199, T1078):** Malware, ransomware, and data breaches exploit various techniques within airport systems. Initial Access tactics, such as Content Injection (T1659) and Exploiting

Public-facing Applications (T1190) enable attackers to infiltrate via public interfaces. Techniques like Drive-by Compromise (T1189), Phishing (T1566), and the use of insecure removable media (T1091) increase the risk of malware and ransomware, ultimately leading to data breaches. External Remote Services (T1133) and Valid Accounts (T1078) allow attackers to leverage stolen credentials to penetrate into the networks, facilitating ransomware and data theft. Risks from Supply Chain Compromise (T1195) and Trusted Relationship (T1199) may introduce vulnerabilities into overall security.

**Supply Chain and Third Party (T1195, T1199):** With multiple party involved, techniques such as Supply Chain Compromise (T1195) and Trusted Relationship (T1199) presents significant threat to airports, allowing attackers to exploit vulnerabilities without raising immediate suspicion.

**Insider Threat (T1091, T1199, T1078):** Techniques such as Insecure Removable Media (T1091) can enable employees to introduce malware into the system. Exploitation of Trusted Relationships (T1199) may allow insiders to manipulate external connections, leading to unauthorized sharing of sensitive information. Additionally, Valid Accounts (T1078) could allow insiders to misuse their credentials.

### 6.4 MITRE Impact (TA0040) for Denial-of-Service

**Denial-of-Service (T1498):** According to the MITRE framework, DoS attacks fall under the Impact tactic, specifically technique ID T1498. This can be executed through methods such as direct network floods (T1498.001) or reflection amplification (T1498.002).

## 7. Modern Defenses for Airport Security

In this section, we outline modern security defense strategies and best practices specifically designed for airport. The key focus areas include the adoption of Cybersecurity Frameworks and the implementation of Zero Trust Architecture, along with additional defenses addressing the identified risks in previous sections.

### 7.1 Cybersecurity Frameworks and Requirements

Numerous studies emphasize the important of adopting cybersecurity frameworks to enhance airport security [1], [5], [8], [41], [55], [67], [68]. Frameworks, such as NIST Cybersecurity Framework, can help identify weaknesses and facilitate development of security objectives. The Civil Air Navigation Services Organization (CANSO) has proposed guidelines to elevate security levels through these frameworks [102], and several airports, including Airports of Thailand, have already implemented such policies [103].

Recently, the TSA issued new cybersecurity requirements, mandating all U.S. airports and aircraft operators to develop cybersecurity policies [104], [105]. Additionally, ICAO has created Standards and Recommended Practices [106], [107] that urge airports to implement cybersecurity risk management frameworks and collaborate to advance ICAO's cybersecurity framework.

### 7.2 Zero Trust Architecture

Zero Trust is a modern cybersecurity framework that prioritizes verifying and protecting all entities based on the principle of least privilege. It involves capturing and analyzing logs for effective threat response, acknowledging that internal threats may stem from untrusted devices and personnel. The following discussion highlights the benefits of implementing Zero Trust architecture in airport security.

**Visibility for Subsystems.** The initial phase of an integrated Zero Trust architecture involves identifying organizational assets, their value, stakeholders, and connectivity. This foundational step prioritizes Zero Trust security policies, including secure access, the principle of least privilege, and enhanced visibility into subsystems.

**Network Segmentation.** Network segmentation is a foundational element of the Zero Trust architecture, allowing network administrators to enforce the principle of least privilege. For instance, the IoT network is isolated from other segments to prevent unauthorized parties from exploiting IoT vulnerabilities and mitigate the risk of lateral movement within the airport's infrastructure.

**Demilitarized Zone.** Internet-based services are prime targets for cyberattacks, making it essential to configure a separate network segment called a "Demilitarized Zone" (DMZ). The DMZ acts as a controlled gateway between the internal network and the internet, enforcing strict connectivity rules through firewalls and packet filtering.

The External Policy Enforcement Point (PEP) in Figure 2 filters malicious internet traffic, while the Internal PEP manages the traffic between DMZ servers and the internal network. This layered security approach ensures that external services remain isolated from internal systems.

**7.3 Security Awareness Training.** The aviation sector may soon face regulatory mandates requiring security awareness training for employees [4], [7]. To effectively implement such programs, organizations actively engage in the development process, providing continuous feedback to enhance the training effectiveness.

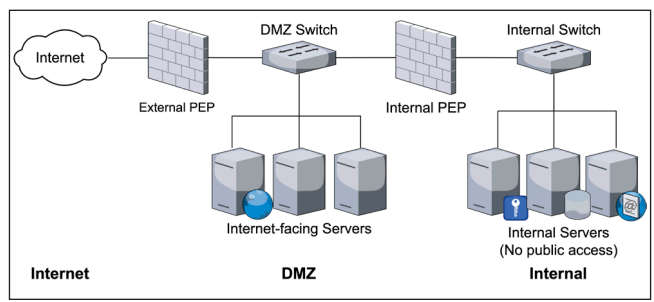

*Figure 2.* DMZ subnet that separates an enterprise internal network from other untrusted networks e.g., the Internet.

**7.4 Malware, Ransomware and Data Breach**

In addition to previously discussed security measures, airports should implement targeted strategies to detect malware and ransomware. The following mitigation strategies can strengthen an airport's cybersecurity posture:

**Endpoint Detection and Response (EDR).** To bolster cybersecurity, airports should implement advanced anomaly detection and monitoring to swiftly identify unusual patterns indicative of ransomware. EDR solutions provide real-time analysis of host activities, enabling rapid detection of malicious behaviors [108].

**Data Backup and Disaster Recovery.** It is imperative to implement comprehensive IT disaster recovery plans that encompass regular off-site data backups. These backups must be secured against common threats, such as ransomware that seeks to compromise backup files (T1486) [109]. Routine testing of backup restoration procedures is crucial to ensure their usability in the event of an incident.

**Data Breach Prevention** involves proactive monitoring of data for irregular patterns—such as unexpected changes in data size, unusual timestamps, or unauthorized access attempts—is essential for early detection of potential breaches. Strong authentication and encryption for data access further safeguard sensitive information from unauthorized users and ensure compliance with relevant regulatory data security and privacy requirements.

**7.5 Supply Chain and Third Party Risk Management**

**Due Diligence in Supply Chain Management.** Any third-party solutions integrated into airport systems must be treated as part of the airport's threat landscape. A rigorous selection process should assess adaptability, security features, secure update mechanisms, and support systems to effectively manage security liabilities and reduce the risk of unforeseen incidents.

**Physical Access Restrictions and Tamper Proofing.** IoT devices deployed throughout airports are susceptible to physical access attacks, where criminals may steal them for unauthorized entry. To mitigate this risk, various tamper-proof techniques can be applied [110], [111]. For instance, devices can be housed in secure or tamper-resistant cases and disabling or factory resetting.

**7.6 Insider Threat Mitigation**

Mitigating insider threats is challenging, as they often bypass traditional security measures. Prevention relies on the principle of least privilege, restricting user access to essential functions, along with monitoring for anomalous behaviors. Implementing a Zero Trust Architecture strengthens security by requiring identity verification for every access request and ensuring all access is logged and analyzed.

**7.7 Denial-of-Service Mitigation**

**Cloud or Hybrid DoS Scrubbing Platforms** enhance security by redirecting traffic through specialized infrastructure that filters out malicious traffic before it reaches the airport's network [112], [113]. Incorporating redundancy and failover mechanisms is also essential, as it improves resilience and minimizes downtime, ensuring essential services remain operational during an attack.

### 7.8 Collaborative Threat Intelligence Sharing

Numerous studies underscore the importance of information sharing within the industry [5], [61], [64], [114]. A real-world example demonstrates the effectiveness: during a spear phishing campaign targeting airport executives, emails containing malware were detected. Through collaborative efforts, the attack was neutralized across the sector [115]. By adopting a multi-faceted approach, including collaborative threat intelligence sharing, airports can enhance their defenses against evolving cyber threats.

## 8. Challenges and Future Research Directions

While a range of defenses have been detailed, significant challenges remain that must be addressed in order to further enhance airport cybersecurity

**Evolving Threat Landscape.** Cyber threats are continuously evolving and becoming more sophisticated, and diversified. This necessitates ongoing research and airport adaptability to counter new attack vectors.

**Resource Constraints.** Smaller airports often face significant resource limitations e.g., budget and personnel, hindering the implementation of cybersecurity measures.

**Integration of Legacy Systems.** Integrating modern security measures with outdated systems presents significant challenges and often requires substantial investment.

**Future Research Directions.** Our literature review reveals a pressing need for further research in key areas: (1) Supply Chain and Third-Party Risks, (2) Cybersecurity Awareness, and (3) Development of Airport-Specific Cybersecurity Frameworks. The limited publications in these domains highlight the unique challenges airports face.

Additionally, future research should investigate the integration of advanced technologies like Machine Learning, AI, and Generative AI, focusing on their effectiveness in enhancing airport operations while mitigating potential vulnerabilities. While existing studies have started to address these gaps, ongoing research is essential to keep up with emerging threats and solutions.

## 9. Conclusion

This paper explores the critical area of airport cybersecurity, highlighting the seriousness of emerging threats in this domain. Through insights gained from recent real-world incidents and a systematic literature review, we conducted a comprehensive analysis and categorized major cybersecurity risks confronting airports, aligned with the MITRE ATT&CK Matrix, providing a valuable framework for exploring practical defenses and best practices articulated in the MITRE knowledge base.

In conclusion, we advocate for the adoption of modern security policies, including robust Cybersecurity Frameworks and Zero Trust Architecture, alongside critical security measures. This study aims to enhance the aviation industry's understanding of the current threat landscape and provide a foundation for enhancing cybersecurity defense and resilience.